\documentclass[a4paper,12pt,nofootinbib]{article}
\usepackage[english]{babel} 
\usepackage[T1]{fontenc}
\usepackage{float}
\usepackage{setspace}
\usepackage{appendix}
\usepackage{parcolumns}
\usepackage{cite}
\usepackage{times}
\usepackage[OT1]{fontenc}
\usepackage{type1cm}
\usepackage{url}
\usepackage{subfigure}
\usepackage{braket}
\usepackage{graphicx}
\usepackage{sidecap}
\usepackage{xspace}
\usepackage{tikz}
\usepackage{blkarray}
\usepackage{indentfirst}
\usepackage[mathscr]{eucal}
\usepackage{geometry}
\usepackage{booktabs}
\usepackage{indentfirst}
\usepackage{bm}
\usepackage{multirow}
\usepackage{dcolumn}
\usepackage{graphicx}
\usepackage{mathrsfs}  
\usepackage{enumerate}
\usepackage[utf8]{inputenc}
\usepackage{amsmath}
\usepackage{amssymb}
\usepackage{amsfonts}
\usepackage{amsthm}
\usepackage{mathrsfs}
\usepackage{graphicx}
\usepackage{multirow}
\usepackage{ulem}	
\usepackage{amsmath,mathtools}
\usepackage{comment}
\usepackage{amsbsy}
\usepackage{hyperref}
\usepackage{bookmark}

\usepackage{nccmath}   

\makeatletter
\@addtoreset{equation}{section}

\makeatother

\def\det{{\rm det}}

\newcommand{\be}{\begin{equation}}
\newcommand{\ee}{\end{equation}}
\newcommand{\bea}{\begin{eqnarray}}
\newcommand{\eea}{\end{eqnarray}}
\newcommand{\vs}[1]{\vspace{#1 mm}}

\newcommand{\nn}{\nonumber\\}
\newcommand{\lc}{{\mit\Gamma}}

\def\cL{{\cal L}}

\def\cP{{\cal P}}

\def\cO{{\cal O}}

\newcommand{\dd}{\mathrm{d}}

\newcommand{\trT}[2]{\mathrm{tr}_{(#1#2)}T}
\newcommand{\divT}[1]{\mathrm{div}_{(#1)}T}

\newcommand{\trdivT}[1]{\mathrm{tr}\mathrm{div}_{(#1)}T}
\newcommand{\DivT}[1]{\mathrm{Div}_{(#1)}T}

\definecolor{monza}{HTML}{CF000F}
\definecolor{bostonuniversityred}{rgb}{0.8, 0.0, 0.0}

\usepackage{xcolor}

\newcommand{\bae}[1]{\begin{align} #1 \end{align}}

\newcommand{\diag}{\mathrm{diag}}

\newcommand{\MTT}{M^{TT}}
\newcommand{\HTT}{H^{TT}}
\newcommand{\HRT}{H^{RT}}
\usepackage{physics}
\newcommand{\calL}{\mathcal{L}}
\usepackage{hyperref}
\hypersetup{backref=true,       
pagebackref=true,               
hyperindex=true,                
colorlinks=true,                
breaklinks=true,                
urlcolor= bostonuniversityred,                
linkcolor= bostonuniversityred,                
bookmarks=true,                 
bookmarksopen=false,
citecolor=bostonuniversityred,
linkcolor=bostonuniversityred,
filecolor=bostonuniversityred,
citecolor=bostonuniversityred,
linkbordercolor=bostonuniversityred
}
\newcommand{\rpm}{m_\mathrm{P}}

\begin{document}

\vs{3}
\begin{center}
{\large\bf Some simple theories of gravity with propagating torsion}
\vs{8}

{\large
Y. Mikura${}^{\dagger *}$\footnote{e-mail address: mikura.yusuke.s8@s.mail.nagoya-u.ac.jp}
\quad
V. Naso${}^*$\footnote{e-mail address: vnaso@sissa.it}
\quad
R. Percacci${}^*$\footnote{e-mail address: percacci@sissa.it}
}
\medskip

${}^\dagger${Department of Physics, Nagoya University,\\
Furo-cho Chikusa-ku, Nagoya, Aichi 464-8602, Japan}\\
${}^*${International School for Advanced Studies, via Bonomea 265, I-34136 Trieste, Italy}\\
${}^*${INFN, Sezione di Trieste, Italy}
\end{center}
\bigskip

{\narrower
{\bf Abstract.}
We consider antisymmetric Metric-Affine Theories of Gravity with a Lagrangian containing the most general terms up to dimension four and search for theories that are ghost- and tachyon-free when expanded around flat space. We find new examples that propagate only the graviton and one other massive degree of freedom of spin zero, one or two. These models require terms of the form $(\nabla T)^2$ in the Lagrangian, that have been largely ignored in the literature.
}

\section{Introduction}

Metric-Affine theories of Gravity (MAGs) are a vast class of extensions
of General Relativity (GR).
A more restricted subclass are the theories with torsion but with metric-compatible connection.
We call these antisymmetric MAGs,
in contrast to symmetric MAGs that have no torsion
but metric-incompatible connection,
or general MAGs that have both torsion and nonmetricity.
Within the subclass of antisymmetric MAGs,
by far the most studied ones are
the so-called Poincar\'e Gauge Theories (PGTs).
These are usually formulated in the tetrad formalism,
since the tetrad is viewed as a gauge field for translations,
and torsion as its curvature.
The tetrad is then joined to the Lorentz gauge field
to form a gauge field for the Poincar\'e group.
The natural Lagrangian for such theories is a Yang-Mills-type Lagrangian
for the Poincar\'e group, which consists of terms
quadratic in the Lorentz curvature and in torsion.
To this, one may add a Palatini term 
(linear in the Lorentz curvature),
so the structure of the Lagrangian is, schematically
\be
\cL_{\mathrm{PGT}}=\rpm^2 F+a T^2+c F^2\ .
\label{PGTact}
\ee
Irrespective of their geometrical interpretation,
we shall call PGT any antisymmetric MAG that has an action
of the form (\ref{PGTact}).

Since the Poincar\'e group (or for that matter already the Lorentz group)
is non-compact, one cannot expect the kinetic term to be positive for
all the components of the field.
In fact, such theories will generally have ghosts or tachyons
when expanded around Minkowski spacetime.
There are, however, particular subclasses of theories that are free of these pathologies.
One of the earliest examples was given by Neville \cite{neville}.
In the late '70s, Sezgin and Van Nieuwenhuizen classified all PGTs that are free of ghosts and tachyons, and are free of accidental gauge symmetries~\cite{Sezgin:1979zf,sezgin2}.~\footnote{An accidental gauge symmetry is a gauge symmetry of the linearized theory that is not a symmetry of the full theory. Such symmetries are generally undesirable.}
Much later, many other cases that also have accidental symmetries
have been classified in \cite{Lin:2018awc} and this work was extended 
to symmetric MAGs \cite{Percacci:2020ddy,Marzo:2021esg,Marzo:2021iok,Baldazzi:2021kaf}.

If we look at these theories as (classical limits of) quantum field theories,
the restriction to Lagrangians of the form (\ref{PGTact}) is quite unnatural. 
In fact, in addition to the terms of the form $F^2$,
there are terms of the form $FDT$, $(D T)^2$, $FT^2$, $T^2DT$ and $T^4$
that have the same dimension four.
We observe that, while the last three types of terms only give cubic and quartic vertices
when expanded around Minkowski space,
the first two affect the two-point function
and therefore are relevant to the issue of a good propagation.
One would expect the Lagrangians~(\ref{PGTact}) not to be closed under renormalization, and this has indeed been shown recently \cite{Melichev:2023lwj}.
Also from the point of view of effective field theories, one expects all terms of dimension two and four to be present in the low energy dynamics of the theory. The issue of having a healthy propagation should therefore be reexamined in the more general class of antisymmetric MAGs with Lagrangians containing general dimension-two and four terms
\bae{
\cL_{\mathrm{C}} = \rpm^2 F+a T^2+c^{FF} F^2+c^{FT}F DT+c^{TT}(D T)^2
\nonumber\\
+c^{FTT}FTT+c^{TTT}TT D T+c^{TTTT}TTTT\ .
\label{aMAGactC}
}
Some such cases have been discussed previously in \cite{Aoki:2019snr}.
In the present paper we continue the search, in a somewhat different direction
(the difference will be explained in more detail in the beginning of Section~\ref{Sec. new MAGs}).
We will not give an exhaustive classification of all ghost- and tachyon-free
antisymmetric MAGs, but rather exhibit some simple examples,
where simple means that they propagate only the usual massless graviton
and one other massive degree of freedom with spin zero, one or two.
In this we follow the constructive strategy outlined in \cite{Baldazzi:2021kaf}:
instead of trying to solve the ghost- and tachyon-freedom conditions
in general, we postulate from the beginning what
degrees of freedom we want to propagate beyond the graviton,
and constrain accordingly the coefficients in the Lagrangian.
This construction is made possible by the use of the spin projector formalism,
that we quickly review in Section~\ref{sec:linac}, and is further simplified by recasting the theory in what we call its Einstein form (as opposed to the Cartan form (\ref{aMAGactC})).
The Einstein form of an antisymmetric MAG is the reformulation where one
changes variables from metric (or tetrad) and connection,
to metric (or tetrad) and contortion:
\be
A=\lc+K ~.
\ee
Here $A$ is the independent dynamical connection,
$\lc$ is the Levi-Civita connection constructed from the metric (or tetrad),
and $K$ is the contortion tensor which is related to the torsion by
\be
\label{TQphi}
T_{\alpha\beta\gamma} = K_{\alpha\beta\gamma}-K_{\gamma\beta\alpha} ~,
\ee
or conversely
\be
K_{\alpha\beta\gamma} = \frac{1}{2}\left(T_{\alpha\beta\gamma}+T_{\beta\alpha\gamma}-T_{\alpha\gamma\beta}\right)\ .
\label{emanuele}
\ee
If we call 
\be
F_{\rho\sigma}{}^\mu{}_\nu=
\partial_\rho A_\sigma{}^\mu{}_\nu
-\partial_\sigma A_\rho{}^\mu{}_\nu
+A_\rho{}^\mu{}_\lambda A_\sigma{}^\lambda{}_\nu
-A_\sigma{}^\mu{}_\lambda A_\rho{}^\lambda{}_\nu ~,
\ee
the curvature of the independent connection $A$, and
\be
R_{\rho\sigma}{}^\mu{}_\nu=
\partial_\rho\lc_\sigma{}^\mu{}_\nu
-\partial_\sigma\lc_\rho{}^\mu{}_\nu
+\lc_\rho{}^\mu{}_\lambda\lc_\sigma{}^\lambda{}_\nu
-\lc_\sigma{}^\mu{}_\lambda\lc_\rho{}^\lambda{}_\nu ~,
\ee
the curvature of the Levi-Civita connection (the Riemann tensor), 
then we have the relation
\begin{eqnarray}
F_{\mu\nu}{}^\alpha{}_\beta & = & 
R_{\mu\nu}{}^\alpha{}_\beta
+\nabla_{\mu}K_{\nu \,\,\, \beta}^{\,\,\, \alpha}
-\nabla_{\nu}K_{\mu\,\,\, \beta}^{\,\,\, \alpha}
+K_{\mu\,\,\, \gamma}^{\,\,\, \alpha}K_{\nu \,\,\, \beta}^{\,\,\, \gamma}
-K_{\nu \,\,\, \gamma}^{\,\,\, \alpha}K_{\mu \,\,\, \beta}^{\,\,\, \gamma} ~,
\label{FtoR}
\end{eqnarray}
where $\nabla$ is the Levi-Civita covariant derivative.
Using these relations, the Lagrangian~\eqref{aMAGactC} can be rewritten entirely in terms
of the Riemann curvature, torsion and their covariant derivatives as
\bae{
\cL_{\mathrm{E}} = \rpm^2 R + a T^2+b^{RR} R^2 + b^{RT} R \nabla T + b^{TT} (\nabla T)^2
\nonumber\\
+b^{RTT} RTT+b^{TTT} TT\nabla T+b^{TTTT} TTTT\ .
\label{aMAGactE}
}
This is what we call the Einstein form of the theory:
it looks like a metric theory of gravity coupled to a peculiar
three-index matter field.
As we shall see, the analysis of the propagating degrees of freedom
is much simpler when the theory is presented in this form, rather than the Cartan form (\ref{aMAGactC}).

\section{Lagrangians}
\label{sec:lags}

The first choice we have to make is what variables to use.
This can be further split into two sub-choices:
the choice between metric, tetrad or even more general frames,
and the choice between $A$ or $T$ as independent variables.
The first choice is a choice of gauge.
Gravity is formulated in the tangent bundle $TM$,
whose transitions functions 
have values in the group $GL(4)$, and can therefore be seen as a gauge theory for the linear group.
We are free to choose the gauge so that $GL(4)$ is completely broken -
this corresponds to using natural (coordinate) frames in $TM$
and the metric as a dynamical variable,
the gauge in which $GL(4)$ is broken down to the Lorentz subgroup -
this corresponds to using orthonormal frames in $TM$
and the tetrad as a dynamical variable,
or keeping the whole $GL(4)$ invariance manifest -
this corresponds to using arbitrary frames in $TM$
and both metric and frame as dynamical variables.
Since in PGTs one has a dynamical Lorentz connection,
these theories are often presented in the second of these gauges,
where Lorentz invariance is manifest.
However, there is no physical difference between the theory
presented in different gauges,
so for the sake of reducing as much as possible the number of variables,
we shall work with natural frames and the metric is treated as a dynamical variable.
As for the second choice, throughout this paper
we will use the Einstein form.

The general structure of the Lagrangians has been given in
Equations (\ref{aMAGactC}) and (\ref{aMAGactE}).
We now have to list in detail all the invariants they contain.
Since we are only interested in the propagators,
it is enough to consider terms at most quadratic in $F$ and $T$
(in the Cartan form) or $R$ and $T$ (in the Einstein form).
These are contained in the first lines of (\ref{aMAGactC}) and (\ref{aMAGactE}).
The terms in the second lines, when expanded around flat space,
only give cubic and quartic interactions.
We therefore only need bases of independent invariants
quadratic in the fields.
Such bases have been worked out in \cite{Baldazzi:2021kaf},
and we report here the results for convenience.

\subsection{Dimension-two terms}

In antisymmetric MAGs there are four dimension-two terms. These give the propagator for the graviton and mass terms for the connection $A$.
In the Cartan form, the dimension-two part of the Lagrangian is
\be
\cL^{(2)}_{\mathrm{C}}=-\frac12\left[-\rpm^2 F+\sum_{i=1}^{3}a^{TT}_i M^{TT}_i\right]\ ,
\label{genPalC}
\ee
where $F=F_{\mu\nu}{}^{\mu\nu}$ is the unique scalar
that can be constructed from the curvature
and $\rpm$ is the reduced Planck mass. The unique scalar $F$ will be referred to as the Palatini term.
The other scalars are
\footnote{We write $\mathrm{tr}_{ab}T^\mu$ for the trace
of torsion on the $a$-th and $b$-th index. 
We write $\mathrm{div}_{a}T^{\mu\nu}$ for the divergence
of torsion on the $a$-th index. }
\be
M^{TT}_1=T^{\mu\rho\nu}T_{\mu\rho\nu}\ ,\quad
M^{TT}_2= T^{\mu\rho\nu}T_{\mu\nu\rho}\ ,\quad
M^{TT}_3= \trT12^{\mu} \trT12_\mu ~,
\label{ttqq}
\ee
where $\mathrm{tr}$ is the trace over the given indices, e.g. $\trT12^b=T_a{}^{ab}$.
Going from the Cartan to the Einstein form,
as discussed in the previous subsection,
yields

\be
\cL^{(2)}_{\mathrm{E}}=-\frac12\left[-\rpm^2 R+\sum_{i=1}^{3}m^{TT}_i M^{TT}_i\right] \ ,
\label{genPalETQ}
\ee
where $T$ now has to be treated as an independent variable and
the coefficients $m^{TT}_i$ are related to the $a^{TT}_i$ 
as in (\ref{mtt1}-\ref{mtt3}).

\subsection{Dimension-four terms}

The dimension-four terms give higher-derivative propagators 
for the graviton and normal propagators for the connection $A$
or the torsion.
Listing such terms is more complicated, because there are several of them
and they are related among themselves by integrations by parts and various identities.
It turns out that there are 14 independent dimension-four invariants
quadratic in $F$, $T$ and their derivatives (in the Cartan form)
or $R$, $T$ and their derivatives (in the Einstein form).
There is arbitrariness  in the choice of the basis for such invariants
and the arbitrariness is greater in the Cartan form,
because $F$ has fewer symmetries and one can construct more invariants.
We will only give two such bases, that are selected for the following reasons.
In the Cartan form, for ease of comparison with the PGT Lagrangians,
we keep all six $FF$ terms, four $FDT$ terms and four $(DT)^2$ terms.
In the Einstein form we keep the three $RR$ terms (familiar from
quadratic metric gravity), all nine $(\nabla T)^2$ terms
and two $R\nabla T$ terms.

In the Cartan form of the theory, the quadratic Lagrangian is
\be
\cL_{\mathrm{C}}^{(4)}=-\frac12\left[\sum_i c^{FF}_i L^{FF}_i
+\sum_i c^{FT}_i L^{FT}_i
+\sum_i c^{TT}_i L^{TT}_i
\right]~,
\ee
where
\bea
L^{FF}_1&=&F^{\mu\nu\rho\sigma}F_{\mu\nu\rho\sigma}\ , \quad
L^{FF}_3=F^{\mu\nu\rho\sigma}F_{\rho\sigma\mu\nu} \ ,\quad
L^{FF}_4=F^{\mu\nu\rho\sigma}F_{\mu\rho\nu\sigma}\ ,
\nonumber\\
L^{FF}_7&=&F^{(13)\mu\nu}F^{(13)}_{\mu\nu}\ ,\quad
L^{FF}_8=F^{(13)\mu\nu}F^{(13)}_{\nu\mu}\ ,\quad
L^{FF}_{16}=F^2 ~,
\label{LFF}
\\
L^{FT}_1&=&F^{\mu\nu\rho\sigma}D_\mu T_{\nu\rho\sigma}\,,\quad
L^{FT}_8=F^{(13)\mu\nu} D_\mu\trT12_{\nu} \ ,
\nonumber\\
L^{FT}_9&=&F^{(13)\mu\nu} D_\nu\trT12_{\mu} \ ,\quad
L^{FT}_{13}=F^{(13)\mu\nu}\, \DivT1_{\mu\nu} \ ,
\label{LFT}
\\
L^{TT}_{1}& =& D^\alpha T^{\beta\gamma\delta} D_\alpha T_{\beta\gamma\delta} \ ,\quad
L^{TT}_{2} = D^\alpha T^{\beta\gamma\delta} D_\alpha T_{\beta\delta\gamma} \ ,
\nonumber\\
L^{TT}_{3} &=& D^\alpha \trT12^\beta D_\alpha\trT12_\beta \ ,\quad
L^{TT}_{5} = \DivT1^{\alpha\beta}\DivT1_{\beta\alpha} ~.
\label{LTT}
\eea
Here $D$ is the covariant derivative formed with the independent connection, e.g.
\be
D_\mu V^\nu=\partial_\mu V^\nu+A_\mu{}^\nu{}_\lambda V^\lambda ~,
\ee
and Div denotes the divergence
formed with $D$, e.g. $\DivT3^{\alpha\beta}=D_\gamma T^{\alpha\beta\gamma}$ etc. 
The non-consecutive numbering comes from compatibility with the general bases defined in \cite{Baldazzi:2021kaf}.
In the Einstein form of the theory, the quadratic Lagrangian is
\be
\cL_{\mathrm{E}}^{(4)}=-\frac12\left[\sum_i b^{RR}_i H^{RR}_i
+\sum_i b^{RT}_i H^{RT}_i
+\sum_i b^{TT}_i H^{TT}_i
\right]\ ,
\label{genlag}
\ee
where
\bea
H^{RR}_1&=&R_{\mu\nu\rho\sigma}R^{\mu\nu\rho\sigma}\ ,\quad
H^{RR}_2=R_{\mu\nu}R^{\mu\nu}\ ,\quad
H^{RR}_3=R^2\ ,
\label{RR}
\\
H^{RT}_{3} &=& R^{\beta\gamma} \divT1_{\beta\gamma} \ ,\quad
H^{RT}_{5} = R\,\trdivT1 \ ,
\label{HRT}
\\
H^{TT}_{1} &=& \nabla^\alpha T^{\beta\gamma\delta} \nabla_\alpha T_{\beta\gamma\delta} \ ,\quad
H^{TT}_{2} = \nabla^\alpha T^{\beta\gamma\delta} \nabla_\alpha T_{\beta\delta\gamma} \ ,\quad
H^{TT}_{3} = \nabla^\alpha\trT12^\beta \nabla_\alpha\trT12_\beta \ ,
\nonumber\\
H^{TT}_{4} &=& \divT1^{\alpha\beta}\divT1_{\alpha\beta} \ ,\quad
H^{TT}_{5} = \divT1^{\alpha\beta}\divT1_{\beta\alpha} \ ,
\nonumber\\
H^{TT}_{6} &=& \divT2^{\alpha\beta}\divT2_{\alpha\beta} \ ,\quad
H^{TT}_{7} = \divT1^{\alpha\beta}\divT2_{\alpha\beta} \ ,
\nonumber\\
H^{TT}_{8} &=& \divT2^{\alpha\beta}\nabla_\alpha\trT12_\beta \ ,\quad
H^{TT}_{9} = (\trdivT1)^2 ~.
\label{HTT}
\eea
Here div is the divergence formed with $\nabla$ as
\be
\nabla_\mu V^\nu=\partial_\mu V^\nu+\mit{\Gamma}_\mu{}^\nu{}_\lambda V^\lambda ~.
\ee
The transformation relating the coefficients $b_i$ to the
coefficients $c_i$ is given in Appendix~\ref{sec:app.mapping}.

\subsection{Linearization}
\label{sec:linac}

We linearize the Einstein form of the theory around Minkowski space
\footnote{We use the metric signature $\eta_{\mu\nu}=\diag\, (-1, +1, +1, +1)$.} 
\be
g_{\mu\nu}=\eta_{\mu\nu}\ ,\quad
T_\rho{}^\mu{}_\nu=0\ .
\ee
The metric is expanded as $g_{\mu\nu}=\eta_{\mu\nu}+h_{\mu\nu}$.
Since the VEV of $T$ is zero,
we shall not use a different symbol for its fluctuation
and identify it with $T$.
By Poincar\'e invariance, the linearized Lagrangian takes the form
\be
S^{(2)} = \frac12\int \frac{\dd^4q}{(2\pi)^4}
\,\Phi^T\,\cO\ \Phi \ ,
\label{linC}
\ee
where $\Phi=(T,h)^T$ and, after Fourier transforming,
$\cO$ is constructed only with the metric $\eta_{\mu\nu}$
and with momentum $q^\mu$.

The fields $T$ and $h$ can be decomposed into irreducible representations
of the rotation group (we assume $q^2\not=0$,
and $O(3)$ is the group that leaves $q^\mu$ invariant).
Table (\ref{irreps}) gives the list of such irreducible representations.
For each representation labelled by spin $J$, parity $\cP$, and possibly
an additional index $i$ to distinguish multiple copies of the same irrep,
there is a projector $P_{ii}(J^\cP)$,
and for each pair of representations with the same spin and parity
but different $i$ and $j$ indices,
there is an intertwiner $P_{ij}(J^\cP)$.
These projectors and intertwiners are constructed with the 
longitudinal and transverse projectors on vectors, defined by
\bae{\label{def of L and T}
    L_\mu{}^\nu = \frac{q_\mu q^\nu}{q^2}~, \quad  T_\mu{}^\nu =\delta_\mu^\nu - L_\mu{}^\nu ~.}
They were introduced for two-index tensors in \cite{rivers,barnes,aurilia}
and for antisymmetric three-index tensors by \cite{Sezgin:1979zf}.
The generalization to arbitrary three-index tensors has been given in
\cite{Percacci:2020ddy,schimidt}.
In the following we refer collectively to all the spin projectors and intertwiners
as ``spin projectors''. 
It should be noted that the spin projectors of \cite{Sezgin:1979zf} and 
\cite{Percacci:2020ddy} are suitable for tensors that are antisymmetric
in the last two indices.
Since our dynamical variable is torsion, and it is antisymmetric in the
first and third index, it is convenient to exchange the first and second indices
of the projectors.
The spin projectors satisfy the orthonormality
\be
P_{ij}(J^\cP)P_{k\ell}(J^{\prime\, \cP'})
=\delta_{JJ'}\delta_{\cP \cP'}\delta_{jk}P_{i\ell}(J^\cP)\ ,
\ee
and the completeness
\be
\sum_J\sum_\cP\sum_i P_{ii}(J^\cP)=\mathbb{I} ~.
\ee

\begin{table}[ht]
    \begin{center}
    \renewcommand{\arraystretch}{1.2}
    \begin{tabular}{|c|c|c|}
    \hline
        & $ha$ & $ta$ \\
    \hline
    $TTT$   & $2^-_2$, $1^-_3$ & $0^-$ \\
    \hline
    $TTL+TLT+LTT$   & -  & $1^+_3$  \\
    \hline
    $\frac32 LTT$    & $1^+_2$, & - \\
    \hline
    $TTL+TLT- \frac12 LTT$  & $2^+_3$, $0^+_3$  & -  \\
    \hline
    $TLL+LTL+LLT$     & $1^-_6$   &  -  \\
    \hline
    \end{tabular}
    \quad
    \begin{tabular}{|c|c|}
    \hline
     & $s$ \\
    \hline
    $TT$ & $2^+_4$, $0^+_5$ \\
    \hline
    $TL$  & $1^-_7$  \\
    \hline
    $LL$   &  $0^+_6$  \\
    \hline
    \end{tabular}
    \end{center}
    \caption{$SO(3)$ spin content of projection operators for torsion (left)
    and graviton (right) ($ta$=totally antisymmetric, $ha$=hook antisymmetric). 
The symbols $L$ and $T$ in the first column refer to the longitudinal
and transverse projectors defined in~\eqref{def of L and T}.
The subscripts distinguish different instances of the same representation. 
The non-consecutive numbering follows from the conventions of \cite{Baldazzi:2021kaf}. }
    \label{irreps}
    \end{table}
    \smallskip

Using these spin projectors, the linearized action
can be rewritten in the form 
\be
S^{(2)}= \frac12\int \frac{\dd^4q}{(2\pi)^4}\sum_{J Pij}\Phi^T(-q)\cdot
a_{ij}(J^\cP)\,P_{ij}(J^\cP)\cdot \Phi(q)\ ,
\label{linactSP}
\ee
where we have suppressed the indices carried by the fields and by the projectors for notational clarity.
The $a_{ij}(J^\cP)$ are matrices of kinetic coefficients,
carrying all the information about the propagation and mixing
between different degrees of freedom.
Their explicit form for antisymmetric MAGs is given in Appendix \ref{sec:app.coefficients}.
Invariance under diffeomorphisms lowers by one the rank
of the coefficient matrices $a(1^-)$ and $a(0^+)$
(this is because the transformation parameter $\xi_\mu$
can be decomposed as a three-scalar and a three-vector).
This is particularly clear in the Einstein form of the theory,
where diffeomorphism invariance implies
\be
\label{fiamma}
a(1^-)_{i7}=a(1^-)_{7i}=0 ~, \quad
a(0^+)_{i6}=a(0^+)_{6i}=0 ~.
\ee
One can fix the gauge redundancy by simply suppressing these rows and columns.
The remaining degrees of freedom can be identified
as the eigenvectors of the coefficient matrices and the
eigenvalues are the inverse propagators.
They are of the form 
\be
\lambda=-b q^2-m^2 ~,
\ee
where $b$ and $m$ are functions of the coefficients in the Lagrangian.
The signs of $b$ and $m^2$ determine the properties of each degree of freedom.

A degree of freedom for which $b=0$ does not propagate.
There are two ways in which this can happen.
If $m^2\not=0$ the equation of motion says that the degree of freedom
is zero. 
On the other hand if $m^2=0$, we have a gauge degree of freedom.
Shifts of that degree of freedom leave the action invariant.
In this case the coefficient matrices have one more null eigenvalue
in addition to the ones due to diffeomorphisms.
As long as one is only interested in the linear theory, there is no need to look further.
However, in view of the nonlinearity of MAG, one should ask
whether the gauge invariance is a genuine feature of the full action,
or just a property of the linearized action.
In the  latter case one speaks of an 
``accidental symmetry''.
Either way, when there are additional gauge symmetries,
one can fix them by removing additional rows and columns.
Once all the gauge degrees of freedom have been removed,
the remaining invertible submatrices are called $b_{ij}(J^\cP)$
and the propagator of the theory, contracted with external sources~\footnote{Here $\tau_{abc}$ is the hypermomentum tensor and $\sigma_{ab}$ denotes the energy-momentum tensor.}
$J=(\tau_{abc},\sigma_{ab})^T$, is
\bae{
\Pi= -\frac12\int \frac{\dd^4q}{(2\pi)^4}\sum_{J Pij}J^T(-q)\cdot
b^{-1}_{ij}(J^\cP)\,P_{ij}(J^\cP)\cdot J(q) ~.
\label{linactSP}
}

Since signs are all-important in this discussion, 
it is useful to keep in mind the case of a scalar field:
in this case the kinetic coefficient $a$ is the same as the
operator $\cO$ and 
\bae{
a=-q^2-m^2 ~,
}
must have $m^2>0$ in order not to have a tachyon.
The corresponding saturated propagator,
obtained by solving the classical equation of motion for the source
and inserting back in the linearized action, is 
\bae{
\Pi= \frac12\int \frac{\dd^4q}{(2\pi)^4}J(-q)\frac{1}{q^2+m^2}J(q) ~.
}
When the sources carry Lorentz indices, one has to pay attention
to the additional signs due to the way the indices are contracted.
For example, in
\bae{
\Pi= \frac12\int \frac{\dd^4q}{(2\pi)^4}J^{\alpha\beta\gamma}(-q)\frac{1}{q^2+m^2}
P_{\beta\alpha\gamma}{}^{\mu\lambda\nu}
\eta_{\lambda\rho}\eta_{\mu\sigma}\eta_{\nu\tau}
J^{\rho\sigma\tau}(q) ~,
}
whenever one of the indices of the projector $P$ is longitudinal,
the corresponding metric gives $-1$,
because in the rest frame where the four-momentum takes the form $q^\mu=(m, 0, 0, 0)$ the longitudinal projector is equal to 1 only in the time-time $(0,0)$ component,
while the transverse projector is equal to 1 only in the spatial $(i,j)$ components. 
Thus there is an overall minus sign whenever the projector $P$ contains an odd number of 
longitudinal projectors.

\section{New ghost- and tachyon-free MAGs}\label{Sec. new MAGs}

In our search of ghost- and tachyon-free MAGs we will use the
Einstein form of the theory.
It has the great advantage that the graviton is carried entirely by the two-index tensor $h_{\mu\nu}$ and therefore its propagator has the form that is familiar
from metric theories of gravity.
In fact in this form, MAG appears as ordinary metric gravity
coupled to a three-index field (torsion) that could be viewed as ``matter''.
If we include in the Lagrangian all terms of dimension up to four,
the gravitational sector includes the terms $H^{RR}_{1,2,3}$
of quadratic gravity.
This theory is well known to have its own ghost/tachyon problem.
While the final verdict on that issue in the quantum context is still not out
\cite{Mannheim:2006rd,Salvio:2018crh,Anselmi:2018ibi,Anselmi:2018tmf,Donoghue:2019fcb,Holdom:2021hlo},
an interesting alternative solution at the classical level has been proposed
recently in \cite{Aoki:2019snr}, that is very close in spirit to the present work.
It was shown in that paper that letting some torsional degrees of freedom
propagate, it is possible to avoid ghosts and tachyons
even in the presence of the quadratic invariants $H^{RR}_{1,2,3}$.

In the present paper we follow an alternative route and
set their coefficients $b^{RR}_{1,2,3}=0$,
thus reducing the theory to Einstein gravity coupled to torsion,
and avoiding at least one possible source of trouble.
We observe that all the examples of ghost- and tachyon-free PGTs
in \cite{Sezgin:1979zf} are of this type.
In principle we could allow $b^{RR}_3\not=0$,
since it does not affect the propagation of the spin-two graviton in flat space.
This would give rise to the propagation of another scalar carried by $h_{\mu\nu}$,
but this has been studied extensively in the literature
and since we are mainly interested in new particles coming from torsion,
we shall not discuss this case here.
Thus, in all cases listed below,
the graviton is $h_{\mu\nu}$ and its propagator is
the standard one from General Relativity.

We now look for examples of MAGs propagating only the graviton
and one other massive degree of freedom.
In order to simplify the search we will impose additional
kinematical constraints on torsion,
and assume that it is either totally antisymmetric
or hook antisymmetric (i.e. its totally antisymmetric part vanishes).
These are invariant subspaces under the action of the group $GL(4)$.
The corresponding degrees of freedom can be read from Table \ref{irreps}.
We shall return later to the general antisymmetric MAG.

We will construct our examples of simple MAGs by imposing that
all the degrees of freedom except for the desired ones do not propagate.
In order to avoid having to investigate whether
a particular gauge symmetry is accidental or not,
we will assume that all the non-propagating degrees of freedom
have nonvanishing masses.

\subsection{Totally antisymmetric case}

The totally antisymmetric part of torsion consists of the states
$1^+_3$ and $0^-$.
In the Lagrangian we impose total antisymmetry of torsion as a kinematical constraint.
In this case $M^{TT}_2=-M^{TT}_1$,
$M^{TT}_3=H^{TT}_3=H^{TT}_8=H^{TT}_9=0$,
$H^{TT}_2$ becomes proportional to $H^{TT}_1$, and $H^{TT}_5$, $H^{TT}_6$, $H^{TT}_7$ become proportional to $H^{TT}_4$, and $H^{RT}_3=H^{RT}_5=0$.
Thus we can set
\bae{
m^{TT}_2=0\ ,\quad
b^{TT}_2=b^{TT}_5=b^{TT}_6=b^{TT}_7=0 ~,
}
without loss of generality, and keep only 
$M^{TT}_1$, $H^{TT}_1$, $H^{TT}_4$
as independent invariants.
The only parts of the coefficient matrices to be considered are:
\bae{
b(2^+)&\equiv a(2^+)_{44}= -\frac{1}{4} \rpm^2 \, q^{2}  ~,
\\
b(1^+)&\equiv a(1^+)_{33}= -\left(b^{TT}_1 +\frac13 b^{TT}_4\right)\, q^2- m^{TT}_{1} ~,
\\
b(1^-)&\equiv a(1^-)_{77}=0 ~,
\\
b(0^+)&\equiv a(0^+)_{\{5,5\}}= \frac{1}{2} \rpm^2 \, q^{2} ~,
\\
b(0^-)&\equiv a(0^-)= -b^{TT}_{1}\, q^{2}- m^{TT}_{1} ~.
}
The matrices pertaining to $2^+$, $1^-$ and $0^+$ are exactly as in pure GR,
and the remaining two are the potentially propagating new degrees of freedom.
Since there is only one state for each spin/parity, the analysis is particularly simple.

If we want only $1^+$ to propagate,
we have to set to zero the coefficient of $q^2$ in $b(0^-)$,
which implies $b^{TT}_1=0$.
This leaves us with
\be
b(1^+)= - \frac13 \left(b^{TT}_{4} q^{2} + 3 m^{TT}_{1}\right)\ .
\ee
Now we have to recall that the spin projector $P(1^+)$ is of the form $TTL$, so that the single longitudinal projector gives rise to an additional minus sign.
Thus the conditions for the $1^+$ state to have a healthy propagation are
\bae{
b^{TT}_4 < 0\ ,\quad
m^{TT}_1 < 0\ .
\label{cond1+}
}
On the other hand, if we want only $0^-$ to propagate,
we have to impose $b^{TT}_4=-3b^{TT}_1$.
In this case, given that the spin projector 
$P(0^-)$ is of the form $TTT$,
the conditions for a healthy propagation of $0^-$ are
immediately obvious from the form of $b(0^-)$:
\be
b^{TT}_1>0 ~,\quad
m^{TT}_1>0 ~.
\label{cond0-}
\ee
Finally, if we allow both degrees of freedom to propagate,
it is clear that there is no choice of parameters that will
prevent one of them from being a ghost or tachyon.

\subsubsection{A check: axial vector torsion}\label{sec. axial vector torsion}

There is a simple alternative way to understand these conditions,
that consists of identifying the $1^+_3$ and $0^-$ degrees of freedom
as the transverse and longitudinal components of an axial vector $v^\mu$,
dual to the torsion tensor \cite{Belyaev:1998ax}:
\be
T_{\alpha\beta\gamma}=\eta_{\alpha\beta\gamma\delta}v^\delta ~,
\ee
where $\eta_{\alpha\beta\gamma\delta}$ is the totally antisymmetric tensor.
In such a case the Lagrangian can be rewritten as
\bae{
\cL_\mathrm{E}=(3 b^{TT}_1+ b^{TT}_4)
\nabla_\mu v_\nu\nabla^\mu v^\nu
-b^{TT}_4\nabla_\mu v_\nu\nabla^\nu v^\mu
+3m^{TT}_1v_\mu v^\mu\ .
\label{veclag}
}
Putting
\bae{
b^{TT}_1=0\ ,\quad
b^{TT}_4=-\frac12\ ,\quad
m^{TT}_1=-\frac16m^2\ ,
}
the Lagrangian (\ref{veclag}) becomes the Proca Lagrangian
that correctly describes a massive spin one state,
whereas for $m^{TT}_1=0$
it becomes the Maxwell Lagrangian 
that correctly describes a massless spin-one state.
This confirms the signs of (\ref{cond1+}).
Note that in the massless case we have $b(0^-)=0$,
which is a symptom of the abelian gauge invariance of the theory.
This is a genuine gauge symmetry of the full action,
not an accidental one.

On the other hand, in order to have only the $0^-$ propagating,
we kill the first term in Equation (\ref{veclag}) by setting $b^{TT}_4=-3b^{TT}_1$
and integrate by parts twice, so that the Lagrangian becomes
\bae{
\cL_\mathrm{E}= 3\left(b^{TT}_1(\nabla_\mu v^\mu)^2+m^{TT}_1 v_\mu v^\mu\right)\ .
\label{elisa}
}
If we decompose $v$ in its transverse and longitudinal parts
\bae{
v_\mu=v^T_\mu+\nabla_\mu\psi\ ,
}
the transverse part is seen not to propagate, whereas for the longitudinal one we 
remain with
\be
\cL_\mathrm{E}= 3\psi\Box(b^{TT}_1\Box-m^{TT}_1)\psi\ .
\label{hdl}
\ee
Defining scalars $\psi_1$ and $\psi_2$ through
\bae{
    \psi_1 =\frac{1}{m^{TT}_1}(b^{TT}_1\Box-m^{TT}_1)\psi ~, \quad \psi_2=\frac{1}{m^{TT}_1}b^{TT}_1\Box\psi ~,
}
the Lagrangian can be rewritten as
\be
\cL_\mathrm{E}= -3 m^{TT}_1\psi_1\Box\psi_1
+3\frac{m^{TT}_1}{b^{TT}_1}\psi_2(b^{TT}_1\Box-m^{TT}_1)\psi_2 ~,
\label{elisa2}
\ee
that indicates the presence of two propagating scalars.
However, the massless scalar is a pure gauge degree of freedom, because the Lagrangian~\eqref{hdl} is invariant under the shift
$\psi\to\psi+\xi$ with $\Box\xi=0$ and only $\psi_1$ is affected by the shift.
Thus there is only one massive scalar, and (\ref{cond0-}) are
the correct conditions for it not to be a ghost or tachyon.
See \cite{Hell:2023mph} for a recent discussion of the appearance of (\ref{elisa})
in higher derivative gravity.

\subsubsection{Cartan form}
Finally we can recast these theories in their Cartan form. When torsion is totally antisymmetric and when we are interested only in the free part of the Lagrangian, 
one can choose the basis of dimension-four operators to consist only of curvature squared terms.
For example, the operators
\be
L^{FF}_1\ ,\quad
L^{FF}_3\ ,\quad
L^{FF}_4\ ,\quad
L^{FF}_7\ ,\quad
L^{FF}_{16}\ ,\quad
\label{Cta}
\ee
form such a basis.
Using this basis,  the Lagrangian that correctly propagates a single $1^+$ state in addition to the graviton is 
\bea
\calL_{\mathrm{C}}&=&\frac{1}{2}\rpm^2 F
 -\frac{1}{2}\left(m^{TT}_1-\frac14 \rpm^2 \right)T_{\mu\nu\rho}T^{\mu\nu\rho}
   \nonumber\\
   && -\frac{1}{6}b^{TT}_4( F_{\mu\nu\rho\sigma}F^{\mu\nu\rho\sigma}
-2 F_{\mu\nu\rho\sigma}F^{\rho\sigma\mu\nu}
+2 F_{\mu\nu\rho\sigma}F^{\mu\rho\nu\sigma})
+O((F,T)^3) ~,
\eea
while the Lagrangian that correctly propagates a single $0^-$ state in addition to the graviton is~\footnote{This combination of curvatures already appeared in \cite{Neville:1978bk}.}
\bea
\calL_{\mathrm{C}}&=&\frac{1}{2}\rpm^2 F
 -\frac{1}{2}\left(m^{TT}_1-\frac14 \rpm^2 \right)T_{\mu\nu\rho}T^{\mu\nu\rho}
   \nonumber\\
   && -\frac{1}{3}b^{TT}_1( F_{\mu\nu\rho\sigma}F^{\mu\nu\rho\sigma}
+ F_{\mu\nu\rho\sigma}F^{\rho\sigma\mu\nu}
- 4 F_{\mu\nu\rho\sigma}F^{\mu\rho\nu\sigma})
+O((F,T)^3) ~.
\eea
In view of the presence of Yang-Mills-like terms, the simplicity of these theories is not at all obvious.

\subsection{Hook antisymmetric case}

Having exhausted the discussion of totally antisymmetric MAGs,
we turn to the hook antisymmetric case.
In this case we have to impose
\be
T^{abc}=T^{acb}+T^{bac} \, .
\ee
This identifies some operators in the Lagrangian with others as 
\bae{
    \MTT_2 = \frac{1}{2}\MTT_1 \qc \HTT_2 =  \frac{1}{2}\HTT_1 \qc \HTT_5 = \HTT_4-\HTT_7 \qc \HTT_6 = 2\HTT_7 ~.
}
As a consequence, we can set 
\be
m^{TT}_2=b^{TT}_2=b^{TT}_5=b^{TT}_6=0 ~,
\label{ha}
\ee
without loss of generality and the Lagrangian simplifies to
\bae{
    \calL_\mathrm{E}=-\frac{1}{2} m^{TT}_1 \MTT_1
    -\frac{1}{2}m^{TT}_3 \MTT_3
    -\frac{1}{2}b^{TT}_1 \HTT_1
    -\frac{1}{2}b^{TT}_3 \HTT_3
    -\frac{1}{2}b^{TT}_4 \HTT_4
    \nonumber\\
    -\frac{1}{2}b^{TT}_7 \HTT_7
    -\frac{1}{2}b^{TT}_8 \HTT_8
    -\frac{1}{2}b^{TT}_9 \HTT_9 
    -\frac{1}{2}b^{RT}_3 \HRT_3
    -\frac{1}{2}b^{RT}_5 \HRT_3 ~.
}
We choose the nondegenerate $b$-matrices as
\bae{
b(2^+)= a(2^+)_{\{3,4\},\{3,4\}}\ ,\quad
b(1^-)=a(1^-)_{\{3,6\},\{3,6\}}\ ,\quad
b(0^+)=a(0^+)_{\{3,5\},\{3,5\}}\ .
}
Then, the nondegenerate coefficient matrix for each spin/parity is given by
\bae{
b(2^+)&\!\!=\!\!\pmqty{-m^{TT}_1-\frac{1}{2}(2b^{TT}_1+b^{TT}_4)q^2 & i\frac{b^{RT}_3}{4\sqrt{2}}q^3 \\ -i\frac{b^{RT}_3}{4\sqrt{2}}q^3 & -\frac{m^2_{\mathrm{P}}}{4} q^2}\, ,
    \label{Eq. hook matrix 2p}\\
b(2^-)&\!\!=\!\!- m^{TT}_1 - b^{TT}_1\, q^2\, ,
    \label{Eq. hook matrix 2m} \\
b(1^+)&\!\!=\!\!-m^{TT}_1-\frac{1}{6}(6b^{TT}_1+b^{TT}_4+2b^{TT}_7) \, q^2\, ,
    \label{Eq. hook matrix 1p}\\  
b(1^-)&\!\!=\!\!{\small\pmqty{-m^{TT}_1\!\!-m^{TT}_3\!\!
-(b^{TT}_1+b^{TT}_3)q^2 & -\frac{1}{2\sqrt{2}}[2m^{TT}_3+(2b^{TT}_3+b^{TT}_8)q^2] \\ -\frac{1}{2\sqrt{2}}[2m^{TT}_3+(2b^{TT}_3+b^{TT}_8)q^2]
& -\frac{1}{2}(2m^{TT}_1\!\!+m^{TT}_3)
-\frac{1}{2}(2b^{TT}_1\!\!+b^{TT}_3\!\!+b^{TT}_4\!\!+b^{TT}_7\!\!+b^{TT}_8)q^2}\, ,}
    \label{Eq. hook matrix 1m}\\
b(0^+)&\!\!=\!\!\pmqty{-\frac{1}{2}(2m^{TT}_1+3m^{TT}_3)-\frac{1}{2}(2b^{TT}_1+3b^{TT}_3+b^{TT}_4+3b^{TT}_9)q^2 & i\frac{b^{RT}_3+6 b^{RT}_5}{4\sqrt{2}}q^3 \\ -i\frac{b^{RT}_3+6 b^{RT}_5}{4\sqrt{2}}q^3 & \frac{\rpm^2}{2} q^2}
    \label{Eq. hook matrix 0p}\, .
}

As in the totally-antisymmetric case, we consider a single-state propagation on top of the graviton. 
The determinants of the coefficient matrices for $2^+$~\eqref{Eq. hook matrix 2p} and $0^+$~\eqref{Eq. hook matrix 0p} contain terms of order $q^6$ coming from their
off-diagonal parts. This indicates that the mixing of torsion and graviton fluctuation
would lead to the propagation of an additional degree of freedom.
For our purposes we can therefore simplify the analysis
by setting $b^{RT}_3=b^{RT}_5=0$.
In this way the graviton is completely decoupled from torsion at quadratic level.

\subsubsection{Propagating graviton and massive $2^+$}

To prevent the propagation of $2^-$ we set $b^{TT}_1=0$, and to prevent the propagation of $1^+$ we set $b^{TT}_7=-b^{TT}_4 /2$.
In order for the torsional $2^+$ state to be a propagating massive degree of freedom, 
we must then have $m^{TT}_1\neq 0$ and $b^{TT}_4 \neq 0$, 
which we assume in the following.
Then, we must remove the terms of order $q^4$ and $q^2$ from the
determinant of $b(1^-)$ and the term of order $q^2$ in 
the $(1,1)$-entry of $b(0^+)$.
These conditions admit the solution
\bae{
m^{TT}_3= -\frac{m^{TT}_1}{\left(b^{TT}_4\right)^2}
\left[\left(b^{TT}_4\right)^2+ 2 b^{TT}_4 b^{TT}_8+ 3\left(b^{TT}_8\right)^2\right]\, ,
\nonumber\\
b^{TT}_3=\frac{\left(b^{TT}_8\right)^2}{2b^{TT}_4} 
\qc 
b^{TT}_9=- \frac{2\left(b^{TT}_4\right)^2+3\left(b^{TT}_8\right)^2}{6 b^{TT}_4}\, .
\label{cond2+}}
With this parametrization the coefficient matrix is reduced to
\be
b(2^+) =-\frac{1}{2}\left(b^{TT}_4q^2 + 2m^{TT}_1\right).
\ee
Now recalling that the spin projector $P(2^+)_{33}$ is of type $TTL$,
the propagator has an additional minus sign,
so that the ghost- and tachyon-free conditions are 
\bae{
b^{TT}_4<0 ~, \quad m^{TT}_1<0 ~.
}

As a check, we can now insert (\ref{cond2+}) 
and the other conditions on the coefficients
in the saturated propagator (\ref{linactSP})
and use the explicit form of the spin projectors.
In principle the resulting expression could contain terms up to $q^{-8}$
(with two powers coming from the kinetic coefficients and six from the spin projectors)
but all terms with 8, 6 and 4 inverse powers of $q$ duly cancel, and we remain with
\be
\Pi=\Pi_\mathrm{{grav}}+\Pi_\mathrm{{massive}}+ (\mathrm{rest}) ~,
\label{satpr}
\ee
where
\bea
\Pi _{\mathrm{grav}} \!&=&\! \frac{2}{\rpm^2}\int \frac{\dd^4q}{(2\pi)^4}\, \sigma^{\mu\nu} (-q) \frac{1}{q^2}\, \left(P_s(2^+)_{\mu\nu\rho\lambda} -\frac12 P_s(0^+,ss)_{\mu\nu\rho\lambda} \right) \sigma^{\rho\lambda}(q)
\nn
\!&=&\! \frac{2}{\rpm^2} \int \frac{\dd^4q}{(2\pi)^4}\, \sigma^{\mu\nu} (-q) \frac{1}{q^2}\left( \eta_{\mu\rho}\eta_{\nu\sigma}-\frac12 \eta_{\mu\nu}\eta_{\rho\sigma}\right) \sigma^{\rho\sigma}(q) ~,
\label{m2c}
\eea
is the standard saturated graviton propagator,
\be
\Pi_\mathrm{{massive}} = \frac{1}{2}\int \frac{\dd^4q}{(2\pi)^4}\, J^{\mu\nu} (-q)  \frac{1}{q^2+m^2} P(2^+,m^2){}_{\mu\nu}{}^{\rho\sigma}  J_{\rho\sigma}(q)\ ,
\label{massive2+}
\ee
is the standard saturated propagator of a massive spin-two particle
and the rest is an expression that does not contain any pole in $q$.
In (\ref{massive2+}), the mass of the canonically normalized field is 
$m^2=2m^{TT}_1/b^{TT}_4$,
the source for the massive spin-two is
related to the source of torsion by
\bae{
        J^{\mu\nu}=\sqrt{\frac{2}{-m^{TT}_1}}\nabla_\rho \tau^{\rho\mu\nu}~,
    }
and
\bea
P(2^+,m^2){}_{\mu\nu}{}^{\rho\sigma}
\!&=&\! T(m^2)_{(\mu}{}^{(\rho} T(m^2)_{\nu)}{}^{\sigma)}
-\frac13 T(m^2)_{\mu\nu} T(m^2)^{\rho\sigma} \,,
\eea
with
\be
T_{\mu\nu}\left(m^2 \right)
= \eta_{\mu\nu} + \frac{q_\mu q_\nu}{m^2} \ ,
\label{Ponemass}
\ee
is the usual transverse-traceless projector, put on shell.

\subsubsection{Propagating graviton and massive $2^-$}

It has already been discussed in \cite{Baldazzi:2021kaf}
how to choose the coefficients of a completely general MAG
in such a way that only the graviton and a massive $2^-$ propagate.
Here we show how to arrive at such a model in the restricted
context of a hook antisymmetric MAG.

The conditions that remove all unwanted propagation
are solved by
\bae{
m^{TT}_3 &= \frac{-24 \left(b^{TT}_1\right)^2 + 16 b^{TT}_1 b^{TT}_8 -3\left(b^{TT}_8\right)^2}{4 b^{TT}_1}m^{TT}_1 ~,
\nonumber\\
b^{TT}_3  & =\frac{-8 \left(b^{TT}_1\right)^2 +4b^{TT}_1 b^{TT}_8-\left(b^{TT}_8\right)^2}{4 b^{TT}_1} ~, 
\quad
b^{TT}_4 =b^{TT}_7=-2 b^{TT}_1 ~,
\quad
b^{TT}_9=-b^{TT}_3 ~.
}

With these parameters, the coefficient matrix for the $2^-$ state becomes simply
\bae{
b(2^-) =-b^{TT}_1 q^2-m^{TT}_1\, ,
}
and, since the projector $P(2^-)$ is of type $TTT$,
the ghost- and tachyon-free conditions are
\bae{
m^{TT}_1 >0 \qc b^{TT}_1 > 0 ~.
}
This agrees with the result of \cite{Baldazzi:2021kaf}.
Proceeding as in the previous section we can recast the saturated
propagator in the form (\ref{satpr}), where now
\be
\Pi_\mathrm{{massive}} =  
    \frac{1}{2}
    \int \frac{\dd^4q}{(2\pi)^4}\, J^{\mu\lambda\nu} (-q)  \frac{1}{q^2+m^2} P(2^-,m^2){}_{\lambda\mu\nu\tau\rho\sigma}  J^{\rho\tau\sigma}(q)\ ,
\label{massive2-}
\ee
where $m^2=m^{TT}_1/b^{TT}_1$ and $P(2^-,m^2)$ is the $2^-$
projector put on shell~\footnote{Its explicit expression is given in Appendix D.6 of~\cite{Baldazzi:2021kaf}.}, and
\begin{equation}
	J^{\mu\nu\rho} = \frac{1}{\sqrt{b^{TT}_1}}\tau^{\mu\nu\rho}\ .
	\end{equation}

\subsubsection{Propagating graviton and massive $1^+$}
In order for the $2^+$ and $2^-$ components not to propagate we must have $b^{TT}_1 =b^{TT}_4 =0$, and then for $1^+$ to propagate we must have $b^{TT}_7\not=0$.
The remaining conditions coming from the $1^-$ and $0^+$ sectors
imply for the other parameters:
\bae{
m^{TT}_3 = -\frac{m^{TT}_1}{4\left(b^{TT}_7\right)^2}\left[4 \left(b^{TT}_7\right)^2+ 4 b^{TT}_7 b^{TT}_8+3 \left(b^{TT}_8\right)^2\right] ~, 
\quad
b^{TT}_3=-b^{TT}_9=\frac{\left(b^{TT}_8\right)^2}{4b^{TT}_7}~.
}
Then the kinetic coefficient is
\bae{
b(1^+) =\frac{1}{3}\left(-b^{TT}_7q^2-3 m^{TT}_1\right).
}
Taking into account that $P(1^+)$ contains a single longitudinal projector, the ghost- and tachyon-free conditions are
\bae{
m^{TT}_1<0 ~, \quad b^{TT}_7<0 ~.
}
The saturated propagator is again of the form  (\ref{satpr}), where now
\be
\Pi_\mathrm{{massive}} = \frac{1}{2}\int \frac{\dd^4q}{(2\pi)^4}\, J^{\mu\nu} (-q)  \frac{1}{q^2+m^2} P(1^+,m^2){}_{\mu\nu\rho\sigma}  J^{\rho\sigma}(q)\ ,
\label{massive1+}
\ee
where the mass of the canonically normalized field is $m^2=3m^{TT}_1/b^{TT}_7$
and its source (an antisymmetric two-tensor) is related to the torsion source by
    \bae{
        J^{\mu\nu}=\sqrt{\frac{2}{-3 m^{TT}_1}}\nabla_\rho(\tau^{\rho[\mu\nu]}+\tau^{\mu\rho\nu}) ~,
    }
and the projector is
\be
P(1^+,m^2)^{\mu\nu}{}_{\rho\sigma}=T^{[\mu}{}_{[\rho}(m^2)T^{\nu]}{}_{\sigma]}(m^2)
= \delta{}^{[\rho}{}_{[\mu} \delta {}^{\sigma]}{}_{\nu]}+2\frac{\delta{}^{[\rho}{}_{[\mu}q{}_{\nu]}q{}^{\sigma]}}{m^2} ~.
\ee

\subsubsection{Propagating graviton and massive $1^-$}
In order to stop the propagation of $2^+$, $2^-$, $1^+$ and $0^-$, the coefficients must satisfy the first four of the following conditions:
\bae{
b^{TT}_1 =b^{TT}_4 =b^{TT}_7=0\ ,\quad 
b^{TT}_9=-b^{TT}_3 \ ,\quad
b^{TT}_8=0\ .
\label{onemin}
}
In order to have only one propagating $1^-$, we must set to zero
the coefficient of $q^4$ in $\det[b(1^-)]$,
and this implies the last condition.
There are two $1^-$ representations in the three-index tensor and their coefficient matrix is now
\bae{
b(1^-)=\pmqty{-m^{TT}_1-m^{TT}_3-b^{TT}_3 q^2 & -\frac{1}{\sqrt{2}}\left(m^{TT}_3+b^{TT}_3 q^2\right)\\
-\frac{1}{\sqrt{2}}\left(m^{TT}_3+b^{TT}_3 q^2\right) & \frac{1}{2}\left(-2m^{TT}_1-m^{TT}_3-b^{TT}_3 q^2\right)}\, ,
}
This matrix has eigenvectors: 
{$(-1/\sqrt2, 1)^T$
with eigenvalue $-m^{TT}_1$, and
$(\sqrt2, 1)^T$
with eigenvalue
\be
\frac12\left(-3b^{TT}_3 q^2-2m^{TT}_1-3m^{TT}_3\right)\, .
\label{oneminus}
\ee
It is the latter combination of states that propagates.
Since the projectors $P(1^-)_{33}$ and $P(1^-)_{66}$ have zero and two
longitudinal indices, respectively, there are no additional signs
and the conditions for ghost- and tachyon-freedom are
\bae{\label{Eq. condition 1m}
b^{TT}_3>0\ ,\quad 2m^{TT}_1+3m^{TT}_3>0\ .
}
The saturated propagator for the massive state is
\bae{
\Pi_\mathrm{{massive}} = 
\frac{1}{2}\int \frac{\dd^4q}{(2\pi)^4}\, J^{\mu} (-q)  \frac{1}{q^2+m^2} T(m^2)_{\mu\nu}  J^{\nu}(q)\ ,
\label{massive1-}
}
where the mass of the canonically normalized field is 
$m^2=(2m^{TT}_1+3m^{TT}_3)/3b^{TT}_3$
and its source (a vector) is related to the torsion source by
\bae{
J^{\mu}=\frac{2}{3\sqrt {b^{TT}_3}}\tau^{\mu\rho}{}_\rho ~.
}

\subsubsection{Propagating graviton and massive $0^+$}

Let us finally focus on the $0^+$ state. By taking
\bae{
b^{TT}_1=b^{TT}_3=b^{TT}_4=b^{TT}_7=b^{TT}_8=0 \, ,
}
all modes except for $0^+$ become non-dynamical. 
Then, its kinetic coefficient is
\bae{
a(0^+)_{33} =-\frac12(3 b^{TT}_9q^2+2m^{TT}_1+3 m^{TT}_3 )\, .
}
The projector $P(0^+)_{33}$ has one longitudinal index,
so there is an additional minus sign.
Thus the conditions for ghost- and tachyon-freedom are
\bae{
 b^{TT}_9<0 \ ,\quad  2 m^{TT}_1+3 m^{TT}_3<0\, .
\label{zeroplus}
}
The saturated propagator of the massive scalar is
\be
\Pi_\mathrm{{massive}} = \frac{1}{2}\int \frac{\dd^4q}{(2\pi)^4}\ J(-q) \frac{1}{q^2+m^2} J(q) ~,
\ee
where $m^2=\left(2m^{TT}_1+3m^{TT}_3\right)/3b^{TT}_9$
and the source is
\bae{
        J=\frac{2}{\sqrt{-3(2m^{TT}_1+3m^{TT}_3)}}\nabla_\mu\tau_\lambda{}^{\lambda\mu} ~.
    }

\subsubsection{A check: vector torsion}

Let us consider the special case of vector torsion:
\be
T_{\mu\nu\rho}=v_\mu g_{\nu\rho}-v_\rho g_{\mu\nu}\ .
\ee
Note that in this case the totally antisymmetric part is zero,
so vector torsion is automatically hook antisymmetric.
If we impose this condition as a kinematical constraint,
and also enforce the restrictions (\ref{ha}) on the coefficients,
the Lagrangian becomes
\bae{
    \calL_{\mathrm{E}}= -\frac{9}{2}b^{TT}_3 \nabla_\mu v_\nu \nabla^\mu v^\nu-\frac{9}{2}b^{TT}_9 \left(\nabla_\mu v^\mu\right)^2 -\frac{3}{2}\left(2m^{TT}_1+3m^{TT}_3\right) v_\mu v^\mu ~.
}
We can then distinguish the two subcases when $v$ is transverse or longitudinal.

Imposing $\nabla_\mu v^\mu=0$, $v$ has spin/parity $1^-$ corresponding to the transverse mode. Then, imposing the conditions (\ref{onemin}) so that only $1^-$ propagates, the linearized Lagrangian becomes
\bae{
    \calL_{\mathrm{E}} = v_\mu \eta^{\mu\nu}\left[-\frac{9}{2} b^{TT}_3 q^2 -\frac{3}{2}\left(2m^{TT}_1+3m^{TT}_3\right)\right]v_\nu ~.
}
So this confirms the ghost- and tachyon-free conditions~\eqref{Eq. condition 1m}.
Imposing that $v_\mu=\nabla_\mu \psi$, with $\psi$ a $0^+$ field
(a true scalar) and imposing
the conditions on the coefficients for the propagation of
$0^+$ only, the linearized Lagrangian becomes
\bae{
    \calL_{\mathrm{E}}= -\frac{9}{2} b^{TT}_9 \psi \Box^2 \psi +\frac{9}{2}\left(2m^{TT}_1+3m^{TT}_3\right) \psi \Box \psi ~.
}
This has the same form as (\ref{hdl}) and, treating it in the same way, we find that the only propagating degree of freedom has linearized Lagrangian
\be
\calL_\mathrm{E}= \frac32
\frac{2m^{TT}_1+3m^{TT}_3}{b^{TT}_9}
\psi_2\,(3b^{TT}_9 q^2+2m^{TT}_1+3 m^{TT}_3)\psi_2\ .
\ee
This confirms the conditions (\ref{zeroplus}).

\subsubsection{General antisymmetric MAGs}

In order to simplify the discussion, in the preceding sections we have
assumed that torsion is either totally antisymmetric or hook antisymmetric.
These kinematical restrictions have the effect of reducing the number of
independent terms in the Lagrangian, so the identification of viable
models with the desired properties becomes algebraically simpler.
One may ask whether simple MAGs with a single healthy propagating degree
of freedom, in addition to the graviton, exist also in the context
of general antisymmetric MAGs.
A priori the answer is not obvious, because we are now working in a larger
space of theories (where in principle more solutions could exist)
but we also have more equations, because there are more degrees of freedom
whose propagation we want to inhibit.
It turns out that the first effect prevails, and there are several more solutions
than in the kinematically restricted cases.
These solutions have different $b$- and $m$-coefficients
and correspondingly different masses.
The reader can derive these models by imposing conditions on
the general kinetic coefficients given in Appendix \ref{sec:app.coefficients},
and following the general procedure outlined in the previous sections.
We will not try to systematically list all these cases.

\section{Discussion}

There are only a handful of PGTs without ghosts and tachyons
and without accidental symmetries \cite{Sezgin:1979zf}.
We have seen that when we allow general dimension-four terms
in the Lagrangian, the number of solutions increases considerably.
In this paper we have only considered ``simple'' MAGs,
in the sense that only one degree of freedom propagates in addition
to the graviton, and we found that there are solutions for every
possible choice of the additional degree of freedom.
The cases we have considered are such that all the non-propagating
degrees of freedom have mass terms and therefore must vanish on shell.
One could also consider cases when these masses go to zero,
but then one would have to make sure that the gauge invariances
arising in this way extend to the nonlinear theory.
One can generalize our construction to
cases with two, three or possibly more healthy propagating degrees of freedom, 
and a full classification seems daunting.
Rather than trying to do that, in our next step we will extend the analysis
of this paper to symmetric MAGs, i.e. theories without torsion
but with nonmetric connection \cite{mp}.

One interesting phenomenon that can present itself is that the good propagation
of one degree of freedom implies some pathology for another.
The simplest example of this is the theory of a vector field.
Normally we choose the Lagrangian of a vector field to be of Proca form,
because we require that only spin one propagates,
but suppose we want also spin zero to propagate.
We can achieve this by adding to the Lagrangian a term of the form $\nabla_\mu A^\mu$.
But it is well known that such a theory has pathologies.
In fact we saw in section~\ref{sec. axial vector torsion} that the mass has necessarily a wrong sign
either for $1^-$ or for $0^+$.
One can trace this problem to the fact the $1^-$ is transverse and $0^+$ is longitudinal.
We have seen that the phenomenon occurs also for vector torsion and for
axial vector torsion, for a similar reason.
In fact, we can say that in general such problems will arise when
a given parameter affects the propagation of degrees of freedom with different
numbers of longitudinal and transverse indices.

We have presented our examples of simple MAGs in the Einstein form,
i.e. treating torsion as an independent field and writing the Lagrangian
in terms of the Riemannian curvature and covariant derivative.
In this form MAG looks like a metric theory of gravity coupled to
a peculiar form of matter.
The alternative Cartan form of MAG, where the Lagrangian is written
in terms of the curvature and covariant derivative constructed from
the independent connection, offers an equivalent description that may be preferable from some points of view. While the models we have presented look somewhat trivial
in the Einstein form, in the sense that they are ``just'' GR coupled to
some scalar, vector or tensor matter,
this ``simplicity'' of the models is much less obvious in the Cartan form,
where Yang-Mills-like terms are present in addition to the Palatini one. In view of possible applications of MAG 
to the dark matter (or dark energy) problems 
it is amusing to observe that the equivalence between the two formulations of MAG
suggests that the dichotomy between modifying gravity 
and adding dark matter may be more a semantic issue
than a real physical one.

\bigskip

\leftline{\bf Acknowledgements}
\noindent
RP wishes to thank Ergin Sezgin for many useful discussions.
This work would not have been possible without
extensive use of the free software packages
{\tt xAct}, {\tt xTensor}, {\tt xPert}, {\tt xTras}. YM is supported by JSPS KAKENHI Grants No.~JP22J22254 and JSPS Overseas Challenge Program for Young Researchers.

\vskip2cm
\goodbreak

\break

\begin{appendix}

\section{Coefficients mapping}
\label{sec:app.mapping}

We give here the general transformation between the Lagrangian coefficients in the Cartan basis and in the Einstein basis.

\small
\bae{
a^{TT}_1 &= \frac{1}{4}\left(\rpm^2 + 4 m^{TT}_1 \right) 
\label{mtt1}
\\
a^{TT}_2 &= \frac{1}{2}\left(\rpm^2 + 2 m^{TT}_2 \right) 
\label{mtt2}
\\
a^{TT}_3 &= -\rpm^2  + m^{TT}_3
\label{mtt3}
\\
c^{FF}_1 &= 2 b^{RR}_1 + b^{RR}_2 + 2 b^{RR}_3 - \frac{1}{4} b^{RT}_3 - b^{RT}_5 - \frac{1}{2} b^{TT}_4 - b^{TT}_6 - \frac{1}{2}b^{TT}_8 + \frac{1}{2} b^{TT}_9 
\\
c^{FF}_3 &= -b^{RR}_1 - \frac{3}{2} b^{RR}_2 - 4 b^{RR}_3 + \frac{1}{2} b^{RT}_3 + 2 b^{RT}_5 + 2 b^{TT}_6 + b^{TT}_7 + b^{TT}_8 - b^{TT}_9 
\\
c^{FF}_4 &= \frac{1}{2} \left(2 b^{RR}_2 + 8 b^{RR}_3 - b^{RT}_3 - 4 b^{RT}_5 + 2 b^{TT}_4 - 4 b^{TT}_6 - 4 b^{TT}_7 - 2 b^{TT}_8 + 2 b^{TT}_9 \right) 
\\
c^{FF}_7 &= \frac{1}{2}\left(-2 b^{RR}_2 - 8 b^{RR}_3 + b^{RT}_3 + 4 b^{RT}_5 + 8 b^{TT}_6 + 2 b^{TT}_7 + 2 b^{TT}_8 - 2 b^{TT}_9 \right) 
\\
c^{FF}_8 &= \frac{1}{2}\left(4 b^{RR}_2 + 8 b^{RR}_3 - b^{RT}_3 - 4 b^{RT}_5 - 8 b^{TT}_6 - 2 b^{TT}_7 - 2 b^{TT}_8 + 2 b^{TT}_9 \right) 
\\
c^{FF}_{16} &= b^{RR}_3 
\\
c^{FT}_1 &= \frac{1}{2}\left(-16 b^{RR}_1 - 8 b^{RR}_2 - 16 b^{RR}_3 + 3 b^{RT}_3 + 8 b^{RT}_5 + 8 b^{TT}_6 + 2 b^{TT}_7 + 4 b^{TT}_8 - 4 b^{TT}_9 \right) 
\\
c^{FT}_8 &= \frac{1}{2}\left(b^{RT}_3 + 8 b^{TT}_6 + 2 b^{TT}_7 + 2 b^{TT}_8 \right) 
\\
c^{FT}_9 &= 2 b^{RR}_2 + 8 b^{RR}_3 - \frac{1}{2} b^{RT}_3 - 2 b^{RT}_5 - 4 b^{TT}_6 - b^{TT}_7 - b^{TT}_8 
\\
c^{FT}_{13} &= \frac{1}{2}\left(4 b^{RR}_2 + 16 b^{RR}_3 - b^{RT}_3 - 8 b^{RT}_5 - 8 b^{TT}_6 - 2 b^{TT}_7 - 4 b^{TT}_8 + 4 b^{TT}_9 \right) 
\\
c^{TT}_1 &= \frac{1}{4}\left(4 b^{RR}_1 + b^{RR}_2 - b^{RT}_3 + 4 b^{TT}_1 + 2 b^{TT}_4 \right) 
\\
c^{TT}_2 &= \frac{1}{2}\left(-4 b^{RR}_1 - 3 b^{RR}_2 - 8 b^{RR}_3 + b^{RT}_3 + 4 b^{RT}_5 + 2 b^{TT}_2 + 4 b^{TT}_6 + 2 b^{TT}_7 + 2 b^{TT}_8 - 2 b^{TT}_9 \right) 
\\
c^{TT}_3 &= b^{RR}_2 + 4 b^{RR}_3 - 2 b^{RT}_5 + b^{TT}_3 + b^{TT}_9 
\\
c^{TT}_5 &= 4 b^{RR}_1 + 2 b^{RR}_2 + 4 b^{RR}_3 - b^{RT}_3 - 2 b^{RT}_5 + b^{TT}_5 - 2 b^{TT}_6 - b^{TT}_7 - b^{TT}_8 + b^{TT}_9 
}
\normalsize

\section{Matrix Coefficients}
\label{sec:app.coefficients}

We list here the coefficient matrices appearing in the decomposition of the Lagrangian (\ref{genlag}) of the Einstein form of the theory.
\small 
\bae{
a(2^{+})_{33}  &= - \frac{1}{2} (2 b^{TT}_{1} + b^{TT}_{2} + b^{TT}_{4} + b^{TT}_{5}) \, q^{2} - \frac{1}{2}(2 m^{TT}_{1} +  m^{TT}_{2}) 
\\
a(2^{+})_{34}  &= i \frac{ b^{RT}_{3}}{4 \sqrt{2}} \, q^3
\\
a(2^{+})_{44}  &= -  \frac{1}{4}(4 b^{RR}_{1} + b^{RR}_{2}) \, q^4  - \frac{\rpm^2}{4} \, q^{2} 
\\
a(2^{-})_{22}  &= - \frac{1}{2} (2 b^{TT}_{1} + b^{TT}_{2}) \, q^{2} - \frac{1}{2} (2 m^{TT}_{1} +  m^{TT}_{2}) 
\\
a(1^{+})_{22} &= -\frac{1}{6} (6 b^{TT}_{1} + 3 b^{TT}_{2} + b^{TT}_{4} -  b^{TT}_{5} + 4 b^{TT}_{6} + 2 b^{TT}_{7}) \, q^2 - \frac{1}{2} (2 m^{TT}_{1} + m^{TT}_{2} ) 
\\
a(1^{+})_{23}  &= -\frac{1}{6 \sqrt{2}} (2 b^{TT}_{4} - 2 b^{TT}_{5} - 4 b^{TT}_{6} + b^{TT}_{7}) \, q^2
\\
a(1^{+})_{33}  &= -\frac{1}{3} (3 b^{TT}_{1} - 3 b^{TT}_{2} + b^{TT}_{4} -  b^{TT}_{5} + b^{TT}_{6} -  b^{TT}_{7}) \, q^2 -(m^{TT}_{1} - m^{TT}_{2})
\\
a(1^{-})_{33} & = -\frac{1}{2} (2b^{TT}_{1} + b^{TT}_{2} + 2b^{TT}_{3}) \, q^2 - \frac{1}{2}(2 m^{TT}_{1} + m^{TT}_{2} +2 m^{TT}_{3}) 
\\
a(1^{-})_{36} & = -\frac{1}{2 \sqrt{2}} (2 b^{TT}_{3} + b^{TT}_{8})\, q^2 -  \frac{1}{\sqrt{2}} m^{TT}_{3} 
\\
a(1^{-})_{37} & = 0 
\\
a(1^{-})_{66}  &= -\frac{1}{2} (2 b^{TT}_{1} + b^{TT}_{2} + b^{TT}_{3} + b^{TT}_{4} + 2 b^{TT}_{6} + b^{TT}_{7} + b^{TT}_{8})\, q^2 - \frac{1}{2} (2 m^{TT}_{1} + m^{TT}_{2} + m^{TT}_{3}) 
\\
a(1^{-})_{67}  &= 0 
\\
a(1^{-})_{77}  &= 0 
\\
a(0^{+})_{33} & = - \frac{1}{2} (2 b^{TT}_{1}+ b^{TT}_{2} + 3 b^{TT}_{3} + b^{TT}_{4} + b^{TT}_{5} + 3 b^{TT}_{9})\, q^2
- \frac{1}{2} (2 m^{TT}_{1} + m^{TT}_{2} + 3 m^{TT}_{3}) 
\\
a(0^{+})_{35} & = i \frac{1}{4 \sqrt{2}} (b^{RT}_{3} + 6 b^{RT}_{5}) \, q^3
\\
a(0^{+})_{36} & = 0 
\\
a(0^{+})_{55} & = - (b^{RR}_{1} + b^{RR}_{2} + 3 b^{RR}_{3}) \, q^4  + \frac{\rpm^2}{2}\, q^2
\\
a(0^{+})_{56} & = 0 
\\
a(0^{+})_{66} & = 0 
\\
a(0^{-}) &=- (b^{TT}_{1} -  b^{TT}_{2})\, q^2 -(m^{TT}_{1} - m^{TT}_{2})
}
\normalsize

\end{appendix}

\end{document}